\documentclass[
preprint,
superscriptaddress,
amsmath,
amssymb,
aps,prb]{revtex4-1}

\usepackage{graphicx}
\usepackage{dcolumn}
\usepackage{bm}
\usepackage{xcolor}
\usepackage[mathlines]{lineno}
\usepackage[normalem]{ulem}
\usepackage[
			colorlinks = true,
            linkcolor = blue,
            urlcolor  = blue,
            citecolor = blue,
            anchorcolor = blue
            ]{hyperref}
\newcommand\rxout{\bgroup\markoverwith{\textcolor{red}{\rule[.5ex]{2pt}{.6pt}}}\ULon}

\begin{document}


%


\title{Simultaneous Observation of Topological Edge State and Exceptional Point in an Open and Non-Hermitian System}
\author{Weiwei Zhu}
\affiliation{MOE Key Laboratory of Advanced Micro-Structured Materials, School of Physics Science and Engineering, Tongji University, Shanghai 200092, People's Republic of China}
\affiliation{Shanghai Key Laboratory of Special Artificial Microstructure Materials and Technology, School of Physics Science and Engineering, Tongji University, Shanghai 200092, People's Republic of China}
\author{Xinsheng Fang}
\affiliation{Shanghai Key Laboratory of Special Artificial Microstructure Materials and Technology, School of Physics Science and Engineering, Tongji University, Shanghai 200092, People's Republic of China}
\author{Dongting Li}
\affiliation{Shanghai Key Laboratory of Special Artificial Microstructure Materials and Technology, School of Physics Science and Engineering, Tongji University, Shanghai 200092, People's Republic of China}
\author{Yong Sun}
\affiliation{MOE Key Laboratory of Advanced Micro-Structured Materials, School of Physics Science and Engineering, Tongji University, Shanghai 200092, People's Republic of China}
\affiliation{Shanghai Key Laboratory of Special Artificial Microstructure Materials and Technology, School of Physics Science and Engineering, Tongji University, Shanghai 200092, People's Republic of China}
\author{Yong Li}
\email{yongli@tongji.edu.cn}
\affiliation{Shanghai Key Laboratory of Special Artificial Microstructure Materials and Technology, School of Physics Science and Engineering, Tongji University, Shanghai 200092, People's Republic of China}
\author{Yun Jing}
\email{yjing2@ncsu.edu}
\affiliation{Department of Mechanical and Aerospace Engineering, North Carolina State University, Raleigh, North Carolina 27695, USA}
\author{Hong Chen}
\email{hongchen@tongji.edu.cn}
\affiliation{MOE Key Laboratory of Advanced Micro-Structured Materials, School of Physics Science and Engineering, Tongji University, Shanghai 200092, People's Republic of China}
\affiliation{Shanghai Key Laboratory of Special Artificial Microstructure Materials and Technology, School of Physics Science and Engineering, Tongji University, Shanghai 200092, People's Republic of China}

\begin{abstract}
  This paper reports on the experimental observation of topologically protected edge state and exceptional point in an open and Non-Hermitian system. While the theoretical underpinning is generic to wave physics, the simulations and experiments are performed for an acoustic system whose structure has non-trivial topological properties that can be characterized by the Chern number provided that a synthetic dimension is introduced. Unidirectional reflectionless propagation, a hallmark of  exceptional point, is unambiguously observed in both simulations and experiments.
\end{abstract}

\maketitle
\noindent\textit{Introduction.--}
Exceptional points (EPs), which are branch point singularities associated with the coalescence of eigenvalues and the corresponding eigenvectors, have recently gained substantial attention owing to its intriguing characteristics in parity-time ($\mathcal{PT}$) symmetry systems\cite{EP2}. A flurry of recent activity has demonstrated extraordinary phenomena associated with the EP in $\mathcal{PT}$ symmetry systems, such as loss induced transparency\cite{lossinducedtransparency}, band merging\cite{bandmerging1}, unidirectional invisibility\cite{unidirectionalinvisibility1,unidirectionalinvisibility2,unidirectionalinvisibility3}, and laser mode selectivity\cite{lasermodeselectivity1,lasermodeselectivity2}. In particular, EPs have been identified in classical wave systems, such as optical\cite{optical1,unidirectionalinvisibility2,lasermodeselectivity1,lasermodeselectivity2}, microwave\cite{CPA6} and acoustical systems\cite{unidirectionalinvisibility4,unidirectionalinvisibility5,unidirectionalinvisibility7}. In fact, EP is a broad concept and can be observed in other non-Hermitian systems besides the $\mathcal{PT}$ symmetry system. For instance, EPs have been found in passive (lossy) systems\cite{passive1,passive2,bandmerging2,bandmerging3,enhancedsensitivity1,enhancedsensitivity4,enhancedsensitivity5,unidirectionalreflectionless1,unidirectionalreflectionless2,unidirectionalreflectionless3}, which can be manifested by a series of features such as  unidirectional reflectionless propagation of waves \cite{unidirectionalreflectionless1,unidirectionalreflectionless2,unidirectionalreflectionless3}.

Meanwhile, topological edge states have grown into a burgeoning research area in condensed matter \cite{topologyreview1,topologyreview2} and classical wave physics \cite{PhotonicTopo1,PhotonicTopo2,PhotonicTopo3,PhotonicTopo4,PhotonicTopo5,PhotonicTopo7,PhotonicTopo8,AcousticTopoMeng,AcousticTopo4,AcousticTopo7}. Topological edge states, or topologically protected edge states, possess the ability of enhancing the field intensity\cite{AcousticTopoMeng}  and could give rise to robust one-way propagation\cite{AcousticTopo4,AcousticTopo5}. Topological edge state and EP, however, are largely considered two unrelated topics. This can be possibly attributed to the fact that most topological edge states are obtained in closed (meaning that the topological edge state forms at the interface between the trivial and non-trivial phases) and Hermitian systems, while EPs are generally harbored in open and non-Hermitian systems. Recently, topological edge states in open systems (meaning that the topological edge state forms at the interface between the non-trivial phase and ``conductor") have been theoretically demonstrated by using one-dimensional resonant photonic crystals with modulated far field couplings\cite{RPhotonicTopoIvchenkoPRL}. This provides us the foundation to investigate a variety of open system phenomena pertinent to topological edge states and a possible route for accessing the EP.

In this paper, we report on the observation of topological edge state and EP in an open system with judiciously tailored losses. Firstly, we obtain topologically protected edge states in a comb-like, quasi-periodic acoustic structure with compound unit cells inspired by the commensurate Aubry-Andre-Harper(AAH) model\cite{AAH1,AAH2,AAH3}. The resulting edge states are locked on one side of the system by a modulated phase. When a critical loss $\Gamma_0$ is introduced to the system, the EP is established, giving rise to unidirectional reflectionless propagation. Interestingly, when the edge state is located in the band gap, a remarkable state can be observed where the reflection is zero for the incident wave from one side and almost unity from the other side. To the best of our knowledge, this is the first experiment demonstrating topologically protected edge states in an open and non-Hermitian system, and also the first one to simultaneously achieve the EP and topological edge state. Our work, therefore, builds a non-trivial connection between two seemingly unrelated fields, i.e., topological edge state and exceptional point.

\noindent\textit{Model.--}
\begin{figure}
\includegraphics{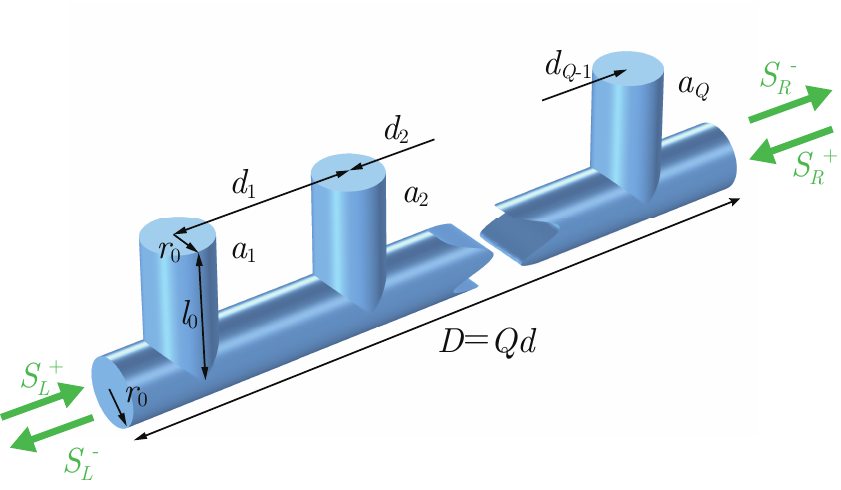}
\caption{\label{fig:1}{\bf The compound unit cell.} The schematic of a comb-like compound unit cell composed of a main tube (radius $r_0$) with $Q$ closed-end, side-branch tubes (radius $r_0$ and length $l_0$). $d_n$ refers to the distance between the $n^{th}$ and $(n+1)^{th}$ side-branch tubes. $a_{n}$ represents the amplitude of the resonant mode of the $n^{th}$ side-branch tube. The dimensions are $r_{0}=14.5$ mm, $d=100$ mm,  and $l_{0}=58$ mm.}
\end{figure}
Inspired by the Aubry-Andr{\'e}-Harper (AAH) model\cite{AAH1,AAH2,AAH3}, a comb-like quasi-periodic acoustic structure with compound unit cells composed of $Q$ side-branch tubes (closed-end tubes) attached to a main tube is constructed [Fig.~\ref{fig:1}]. The distance between the $n^{th}$ and $(n+1)^{th}$ side-branch tubes in the compound unit cell is modulated as
\begin{equation}\label{eq1}
        d_{n}=d[1+\delta\cos(2\pi b(n-1)+\phi)].
\end{equation}
Here $d$ is the unmodulated distance, $\delta$ represents the strength of cosine modulation, $b=P/Q$ is a rational number with $P$ and $Q$ being two integers with no common factors, and $\phi$ defines an arbitrary phase. $\phi$ has an allowed value ranging from $-\pi$ to $\pi$ and is the key parameter to relate a one dimensional system to a two dimensional topologically nontrivial system that can be characterized by the Chern number\cite{AAH1,AAH2,AAH3}.

Since the main tube is of subwavelength size (radius $r_0$), only the fundamental mode (plane wave) is allowed. The acoustic compound unit cell can be described by the temporal coupled mode equation for the lowest resonant mode amplitudes $\tilde{a}_{n}=a_{n}e^{i\omega t}$ of the $n^{th}$ side-branch tube\cite{CMT},
\begin{equation}\label{eq2}
  \frac{d}{dt}\tilde{a}_{n}=(i\omega_{0}-\Gamma)\tilde{a}_{n}-\gamma\sum_{n'}\Lambda_{nn'}\tilde{a}_{n'} +i\sqrt{\gamma}e^{-ik|x_{n}|}\tilde{S}_{L}^{+}+i\sqrt{\gamma}e^{-ik|D-x_{n}|}\tilde{S}_{R}^{+},
\end{equation}
where $\omega_{0}$ is the resonant frequency, $\Gamma$, $\gamma$ are the dissipative and radiative decay rate, respectively, $\Lambda_{nn'}=e^{-ik|x_{n}-x_{n'}|}$ with $x_{n}$ being the location of the side-branch tubes ($d_{n}=x_{n+1}-x_{n}$), $k=\omega/c$ is the wave number in air with speed of sound $c$, and $D=Qd$ is the period of the compound unit cell. $\tilde{S}_{L}^{+}=S_{L}^{+}e^{i\omega t}$ and $\tilde{S}_{R}^{+}=S_{R}^{+}e^{i\omega t}$ correspond to the incident acoustic waves from the left and right ports, respectively. The output acoustic waves $S_{L}^{-}$ and $S_{R}^{-}$ are linked to the input waves by the scattering matrix,
\begin{equation}\label{eq3}
  \left(\begin{array}{c}
    S_{L}^{-} \\
    S_{R}^{-}
  \end{array}\right)=\left(\begin{array}{cc}
                            t &  r_{L} \\
                             r_{R}& t
                           \end{array}
  \right)\left(\begin{array}{c}
                 S_{R}^{+} \\
                 S_{L}^{+}
               \end{array}
  \right).
\end{equation}

Protected by the reciprocity, the transmission coefficients in the cases of left incidence and right incidence must be the same. The transmission comes from the interference of the incident wave and the re-radiated waves from the side-branch tubes. By combining Eqs.~(\ref{eq2}) and (\ref{eq3}), the complex transmission coefficient can be obtained by considering $S_{R}^{+}=0$, and it reads
\begin{equation}\label{eq4}
  t=e^{-ikD}+i\sqrt{\gamma}\sum_{n=1}^{Q}e^{-ik(D-x_{n})}\frac{a_{n}}{S_{L}^{+}}.
\end{equation}
The reflection coefficients in the left incidence and right incidence cases are in general different due to the lack of mirror symmetry of the compound unit cell. The reflection coefficient for the left incidence case can be obtained with $S_{R}^{+}=0$ and it yields
\begin{equation}\label{eq5}
  r_{L}=i\sqrt{\gamma}\sum_{n=1}^{Q}e^{-ikx_{n}}\frac{a_{n}}{S_{L}^{+}},
\end{equation}
while the reflection coefficient with $S_{L}^{+}=0$ in the right incidence case is
\begin{equation}\label{eq6}
  r_{R}=i\sqrt{\gamma}\sum_{n=1}^{Q}e^{-ik(D-x_{n})}\frac{a_{n}}{S_{R}^{+}}.
\end{equation}

\noindent\textit{Band structure.--}
\begin{figure}
\includegraphics{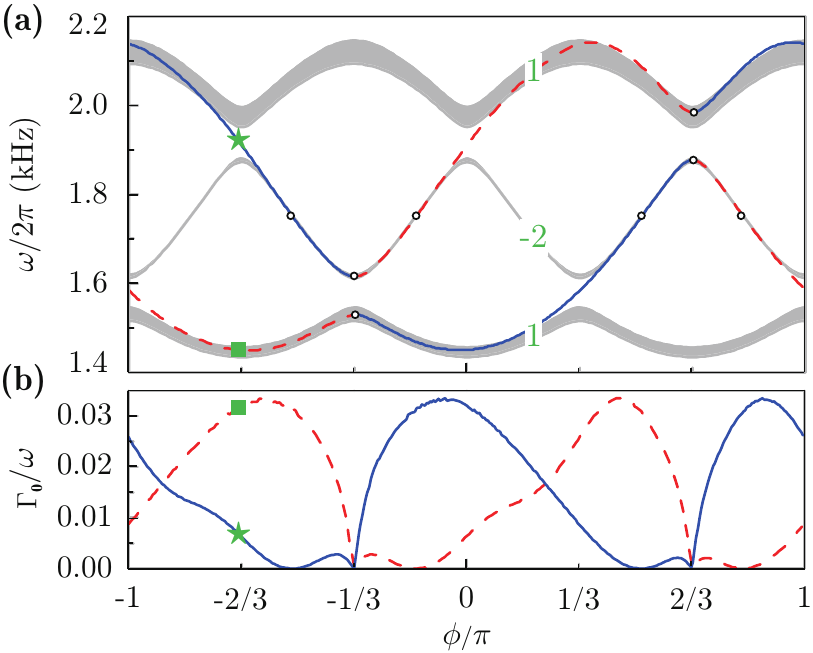}
\caption{\label{fig2}{\bf Topologically protected edge modes.} (a) Projected acoustic band structure of the compound unit cell with $b=1/3$ as a function of $\phi$. The gray regions represent three pass bands separated by two band gaps. The Chern number for each pass band is marked. The eigenfrequencies of the edge modes under left (right) incidence is illustrated with blue (red dashed) line. White circles indicate the points where $\Gamma_0=0$. (b) The critical loss $\Gamma_0$ of the edge modes for left (blue line) and right (red dashed line) incidence. The left edge state and right edge state at $\phi=-2.1$ ($\phi/\pi=-2/3$) are marked by star and square respectively. The parameters for the calculation are $\omega_{0}/2\pi=1.75$ kHz  and $\gamma/2\pi=0.48$ kHz, which are extracted from a full wave simulation.}
\end{figure}
We first investigate the band structure of the acoustic system without dissipative losses, i.e., $\Gamma=0$. The targeted frequency range is determined by the resonance frequency of the side-branch tubes, $\omega_0 \approx c \pi/l_0/2$. In the absence of modulation ($\delta=0$), a flat band would appear at the resonance frequency of the side-branch tubes (see Supplementary Note 1), which can be attributed to the strong interplay between the Bragg scattering (destructive interference of the re-radiated waves from two adjacent side-branch tubes with distane $d=\lambda_{0}/2$ and $\lambda_0=2\pi c/\omega_0$) and local resonance. The introduction of modulation ($\delta\neq0$) forms a compound unit cell with an enlarged period (from $d$ to $D=Qd$). Consequently, the flat band splits into $Q$ bands due to band folding. The band structure of the compound unit cell can be calculated by using the scattering matrix $S_{1}= \left(\begin{array}{cc}
t & r_{L} \\
	r_{R} & t
	\end{array}\right)$ and the Bloch boundary condition $S_{R}^{-}=e^{ik_{B}D}S_{L}^{+}$ and $S_{R}^{+}=e^{ik_{B}D}S_{L}^{-}$,
\begin{equation}\label{eq7}
        \left| {{S_1} - \left( {\begin{array}{*{20}{c}}
{{e^{ - i{k_B}D}}}&0\\
0&{{e^{i{k_B}D}}}
\end{array}} \right)} \right| = 0.
\end{equation}

Figure~\ref{fig2}(a) displays the projected band structure as a function of $\phi$ for a compound unit cell with $\delta=0.4$ and $b=1/3$, indicating that there are three side-branch tubes in a unit cell. The flat band splits into three bands (gray regions) separated by two band gaps [Fig.~\ref{fig2}(a)]. Cosine modulation is employed here to produce topologically nontrivial states, which has been proven to be an effective way both in the tight binding model\cite{AAH1,AAH2,AAH3} and scattering system \cite{RPhotonicTopoIvchenkoPRL}. To confirm the topologically nontrivial nature of our system, the Chern numbers of the three bands are obtained from the phase spectroscopy of the semi-infinite structure\cite{PhaseHafeziPRA} (see Supplementary Note 2), and are marked in Fig.~\ref{fig2}(a). The Chern numbers, i.e., $C={1,-2,1}$, suggest the existence of topological edge modes in the two band gaps according to the bulk-boundary correspondence developed for the Hermitian system\cite{AAH1,AAH2,AAH3}.

\noindent\textit{Topological edge states. --}
For a semi-infinite system, an interesting property of the topological edge mode is that the reflectivity reduces to zero with a proper amount of loss\cite{RPhotonicTopoIvchenkoPRL}. Therefore, the edge state can be retrieved from the zeros of the reflection coefficient, i.e., $r_{\infty}(\omega,\Gamma)=0$, with critical loss $\Gamma=\Gamma_0$. We will later on show that this coincidentally links the edge state to EP. For the current structure, the zeros of $r_{\infty}(\omega,\Gamma)$ are identical to the zeros of $r(\omega,\Gamma)$ for a unit cell (see Supplementary Note 3). The eigenfrequencies of the topological edge modes localized on the left (right) boundary can be calculated under the condition $r_{L}(\omega,\Gamma)=0$ ($r_{R}(\omega,\Gamma)=0$). Edge states exist when the bands possess different Chern numbers. It is not surprising that two edge states exist for our modulated compound unit cell for any given $\phi$: one for the left incidence and one for the right incidence [Fig~\ref{fig2}(a)].

The critical loss $\Gamma_0$ for left edge state and right edge state is plotted as a function of $\phi$, shown in Fig.~\ref{fig2}(b). The points where $\Gamma_0=0$ are marked by circles in Fig.~\ref{fig2}(a). Among these points, there are four of them  at $\phi=-\pi/3$ and $2\pi/3$, where $d_{1}$ equals $d_{2}$, and the eigenmodes of the unit cell have symmetry and anti-symmetry modes (see Supplementary Note 4). The rest of them on the left (right) edge state are located at the resonance frequency of the side-branch tube (around $1750$ Hz) with $d_{1}=d$ ($d_{2}=d$), and the unit cell has a bound state which stems from the destructive interference of the waves from two side-branch tubes. This was recently studied and termed bound states in the continuum\cite{BIC,BIC2}.
\begin{figure}
\includegraphics{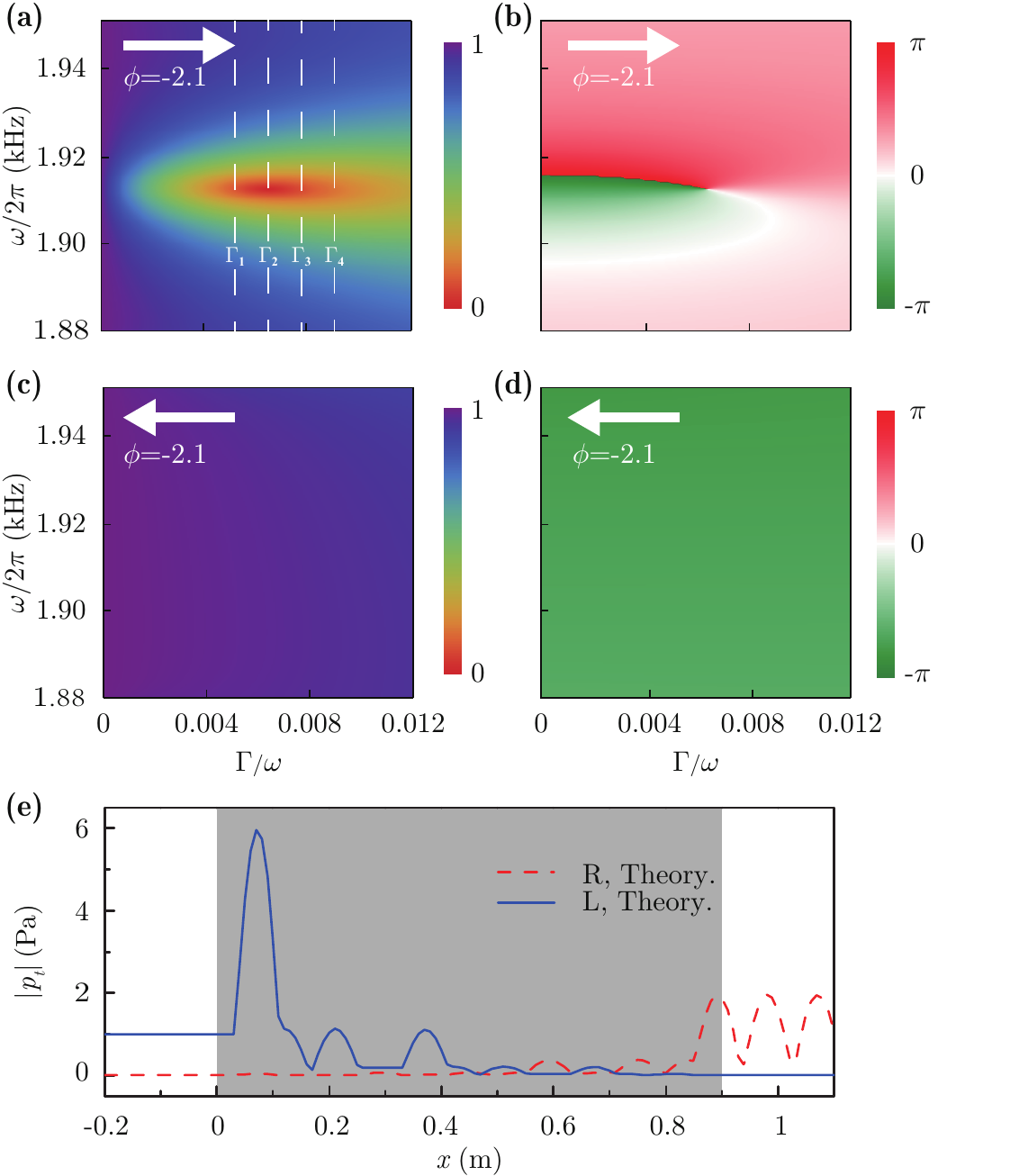}
\caption{\label{fig3}{\bf The calculated reflectivity and reflection phase of the finite structure with three unit cells as a function of $\Gamma$ in the second band gap with $\phi=-2.1$.} (a)(b) for the left incidence and (c)(d) for the right incidence. $\Gamma_{1}/$Re$(\omega)$, $\Gamma_{2}/$Re$(\omega)$, $\Gamma_{3}/$Re$(\omega)$, and $\Gamma_{4}/$Re$(\omega)$ are the loss values used in the simulation and experiment, and they are 0.0051, 0.0063, 0.0074, and 0.0086, respectively.(e) the field distribution at the point ($\omega_0/2\pi,\Gamma_0/\omega_0$) = ($1913 Hz,0.0063$) for left input and right input.}
\end{figure}

\noindent\textit{Topological edge state in a finite structure and the EP. --}
Next, we study the topological edge state in a finite structure composed of $N$ unit cells with $\phi=-2.1$, where a left topological edge state lies in the second band gap marked by the green star in Fig.~\ref{fig2}(a). The transport properties of such system can be described by the scattering matrix $S_{N}=\left(\begin{array}{cc}
t_{N} & r_{L(N)}\\
r_{R(N)} & t_{N}
\end{array}\right)$,
which can be obtained from Eqs.~\ref{eq2}, \ref{eq4}, \ref{eq5}, and \ref{eq6}. The reflectivity of three unit cells for left input is calculated as a function of $\omega$ (in the second band gap frequency range) and $\Gamma$, as shown in Fig.~\ref{fig3}(a). A zero reflectivity is seen at the point ($\omega_0/2\pi,\Gamma_0/\omega_0$) = ($1913 Hz,0.0063$). The pressure field distribution for this point with waves coming from the left side is calculated and shown in Fig.~\ref{fig3}(e). The field is significantly enhanced at the left edge, confirming the existence of the edge state. Furthermore, the eigenvalues $t_{N}\pm\sqrt{r_{L(N)}r_{R(N)}}$ and eigenvectors $\left(\begin{array}{c} 1 \\
\pm\sqrt{r_{R(N)}/r_{L(N)}}
\end{array}\right)$ of the scattering matrix coalesce simultaneously at this point, which is the main characteristic of the EP. In other words, the EP can be simultaneously observed and the reflectivity reduces to zero at the EP. The phase of the reflected wave is also shown in Fig.~\ref{fig3}(b), and a vortex-like pattern centered at the EP is noted. For the edge state locked on the left side at this particular point, extremely asymmetric transport is achieved. The reflectivity and the reflection phase for the right input case shown in Figs.~\ref{fig3}(c)(d) remain almost unchanged in the parameter space. The reflectivity remains almost unity for $\omega$ within the band gap.

\begin{figure}
\includegraphics{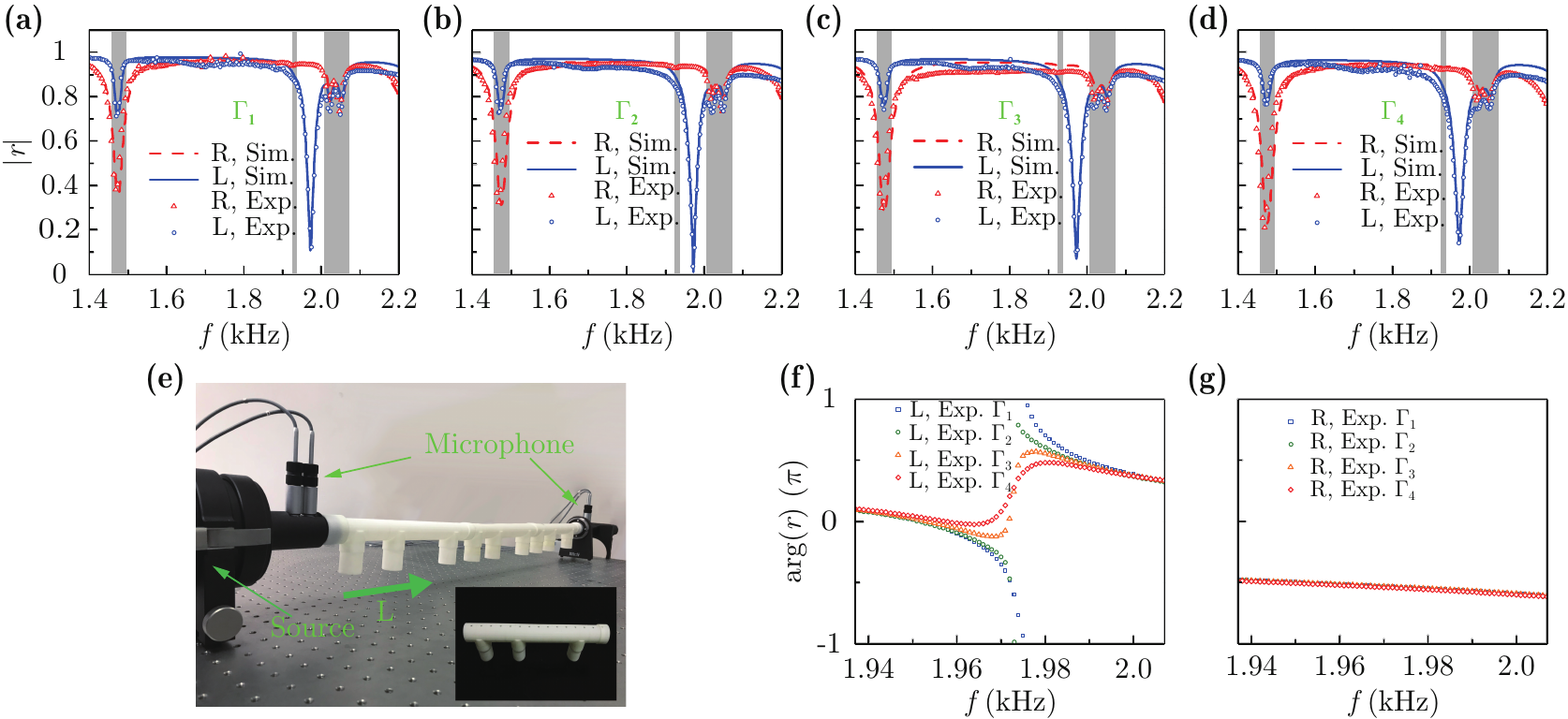}
\caption{\label{fig4}{\bf Simulated and measured reflectivity and reflection phase of the finite structure with three unit cells and different losses $\Gamma_{1}$, $\Gamma_{2}$, $\Gamma_{3}$, and $\Gamma_{4}$.} (a)(b)(c)(d) are the reflectivity for $\Gamma_{1}$, $\Gamma_{2}$, $\Gamma_{3}$, and $\Gamma_{4}$, respectively. The blue solid lines (orange dashed lines) are the simulated results for left incidence (right incidence). The blue open circles (orange circles) are the experimental results for left incidence (right incidence). The three gray regions represent the pass bands. (e) the experimental setup. The inset figure is one unit cell of the sample and the small side holes are used to measure the field distribution. The measured reflection phases at different losses (f) for left incidence and(g) right incidence.}
\end{figure}

To confirm our theory, a series of acoustic experiments are conducted to measure the reflection coefficients of a sample composed of three unit cells with four different loss values corresponding to  $\Gamma_{1}$, $\Gamma_{2}$ (critical loss), $\Gamma_{3}$, and $\Gamma_{4}$ in Fig.~\ref{fig3}(a). The experimental setup for the left incidence case is shown in Fig.~\ref{fig4}(e). For the right incidence case, we simply reverse the sample. The loss is adjusted by placing sponges with $3 mm$ thickness into the side-branch pipes. No sponge is used for $\Gamma_{1}$. In this case, the loss is solely from the intrinsic thermo-viscosity in the tubes. The measured results on the reflectivity are shown in Figs.~\ref{fig4}(a-d) for $\Gamma_{1}$, $\Gamma_{2}$, $\Gamma_{3}$, and $\Gamma_{4}$, respectively. It is seen that the reflectivity for the left input case at $1973$ Hz (this frequency is slightly different from the theoretically predicted one, i.e., 1913 Hz) decreases first from $\Gamma_{1}$ to $\Gamma_{2}$ and then increases from $\Gamma_{2}$ to $\Gamma_{3}$, and to $\Gamma_{4}$. It is also observed that the reflectivity for the right input wave is almost unity in the second band gap. A series of simulations using COMSOL Multiphysics are performed to corroborate the measurement results. These results are shown by lines in Figs.~\ref{fig4}(a-d), where excellent agreements can be observed.
In addition to the left edge state located in the second band gap, a right edge state located in the first pass band also exists when $\phi=-2.1$. As shown in Figs.\ref{fig4}(a-d), the right incidence reflectivity at $1468$ Hz progressively declines as we crank up the loss. In fact, null reflectivity at this frequency is also possible, provided that more loss is introduced to the system as shown in Fig.~\ref{fig2}(b) by the red dashed line (i.e., more loss is needed to balance the leakage loss in this case). However, unlike the left edge state case where the right incidence yields a strong reflection, the left incidence in the right edge state case always has a reflectivity dip at $1468$ Hz since it is in a pass band.

The measured reflection phases in the second band gap for different losses are shown in Fig.~\ref{fig4}(f) for left incidence and in Fig.~\ref{fig4}(g) for right incidence. The left reflection phase has a dramatic transition from $-\pi$ to $\pi$ for a lower loss with $\Gamma_{1}$, which is the characteristic of the Lorentz resonance. When the loss is sufficiently large, which is greater than the critical loss $\Gamma_0$, the reflection phase changes smoothly for the cases of $\Gamma_{3}$ and $\Gamma_{4}$. Meanwhile, the right incidence reflection phase remains almost unchanged with different losses as shown in Fig.~\ref{fig4}(g).
\begin{figure}
\includegraphics{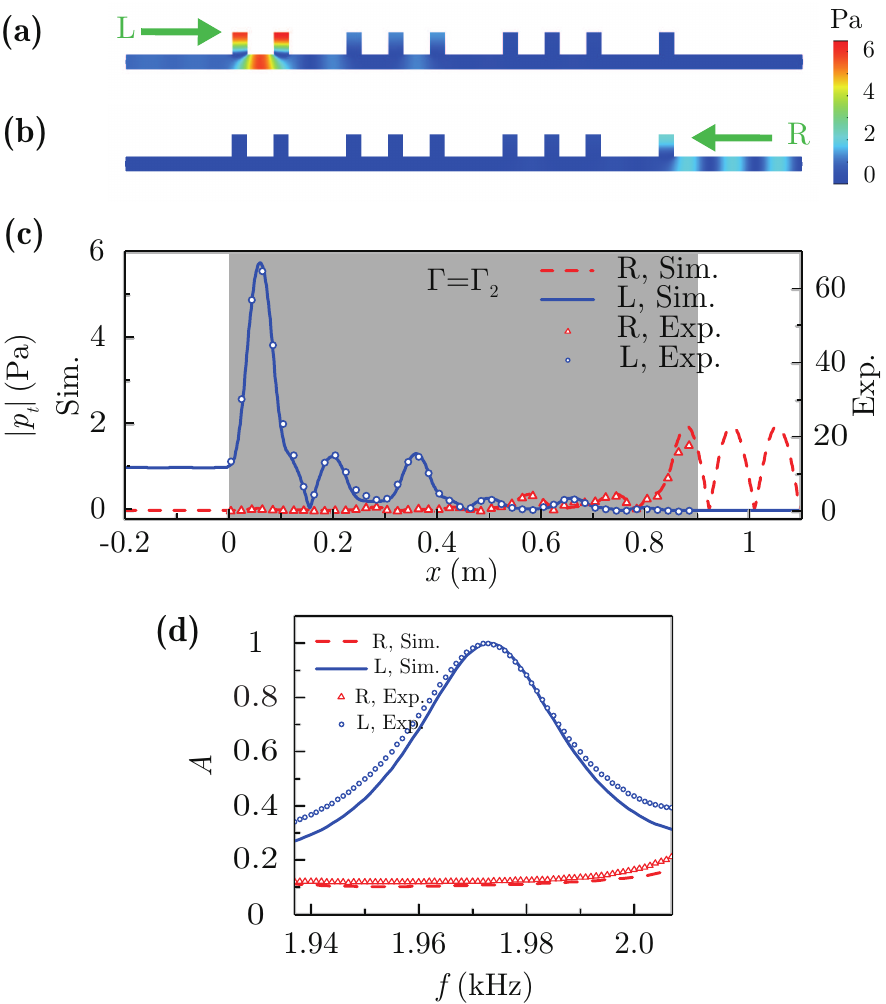}
\caption{\label{fig5}{\bf The pressure field distributions for the left edge state and sound absorption.} (a),(b) Simulated pressure fields for the left edge mode with $\phi=-2.1$ at a frequency of $f=1973$ Hz under the left and right incidence for $\Gamma=\Gamma_2$. (c) Pressure distributions along the main axis of the main tube for the left incidence (blue line) and right incidence (red dash line) cases. (d) Simulated and measured sound absorption at frequencies in the second band gap where $\Gamma=\Gamma_{2}$. Blue solid line (orange dashed line) represents the simulated results for left incidence (right incidence). Blue open circles (orange open triangles) are the experimental results for left incidence (right incidence). Extremely asymmetrical sound absorption is visible at $1973$ Hz. }
\end{figure}


\noindent\textit{Field. --}
Finally, the internal pressure field at $1973$ Hz with loss $\Gamma=\Gamma_{2}$ is also measured. Side holes are drilled on the main tube of the compound unit cell as shown in the inset figure of Fig.~\ref{fig4}(e). For each measurement, only one side hole is used while others are sealed to avoid unnecessary leaking. The measured results are shown in Fig.~\ref{fig5}(c) for both incoming wave directions. The simulated results are shown by the solid line for incoming waves from the left and dashed line for the other direction. The simulated and measured results are in excellent agreement. The field in the interior of the structure is significantly enhanced at the left end for the left incidence case, again confirming the existence of the edge state mode in this open and non-Hermitian system.

\noindent\textit{Unidirectional perfect absorption. --}
The transmissivities for acoustic waves traveling to the left and right are identical due to the reciprocity. However, at the EP, whose frequency is 1973 Hz and is in the bandgap, the reflectivity for the left incidence case can be tuned to zero as illustrated in  Fig.~\ref{fig4}(b) with $\Gamma=\Gamma_{2}$ and the reflectivity for the right incidence case is almost unity (not perfectly unity due to the loss), which means that the sound transmission is near-zero. In other words, this unique state gives rise to perfect absorption for acoustic waves traveling from left to right, and a very strong reflection for waves traveling from right to left. The measured and simulated (COMSOL) sound absorption coefficients defined by $A=1-|t|^{2}-|r|^{2}$ in the second band gap frequency range for $\Gamma=\Gamma_{2}$ are plotted in Fig.~\ref{fig5}(d). The absorption at $1973$ Hz is almost $1.0$ for the left incident wave and is significantly weaker when the incoming wave is from the other direction. This unidirectional phenomenon exists as long as the system has topological edge states located in the band gap and adequate losses are provided.

\noindent\textit{Conclusion. --}
Inspired by the commensurate AAH model, we obtain topologically protected edge states in an open and non-Hermitian system, consisting of a comb-like, quasi-periodic acoustic structure with compound unit cells. The topologically protected edge states are always located at the left edge or right edge of the structure. This enables us to gain asymmetrical transport properties and access EPs by introducing losses into the system. In particular, by choosing the edge state which is located in the band gap, a remarkable state with zero reflectivity from one side and almost total reflection from the other side is obtained. Under the constraint of energy conservation and the reciprocity law, unidirectional perfect sound absorption is observed. The acoustic field measurement affirms the existence of the predicted edge state. Our theory is generic to wave physics and can be readily extended to photonics and plasmonics. Our work, therefore, provides a new route to studying the connection between topological edge state, non-Hermitian physics, and exceptional point.

\noindent {\bf{Method}} \\
\noindent \textbf{Simulations.}
All simulation results are obtained from the acoustics module of the COMSOL Multiphysics Version 5.3 with the air speed of sound $c=343$ m/s and mass density $\rho=1.3$ kg/m$^3$. The loss was considered by the imaginary part of the speed of sound. \\
\noindent \textbf{Sample fabrication and experimental setup.}
We fabricated samples by using the 3D-printing technology, via laser sintering stereo-lithography (SLA, 140 microns) and with photosensitive resin (UV curable resin), where the nominal manufacturing precision is $0.1$ mm. The measurements of the pressure field were performed using a 1/8-inch condenser microphones. The reflection, transmission, and absorption coefficients were measured using a commercial impedance tube (Br\"{u}el \& Kj{\ae}r type-4206T) with a diameter of $29$ mm, four 1/4-inch condenser microphones (Br\"{u}el \& Kj{\ae}r type-4187) situated at designated positions to sense the amplitude and phase of local pressure. During the experiment, the digital signal generated by the computer was sent to the power amplifier (Br\"{u}el \& Kj{\ae}r) and then powered the loudspeaker.

%

\noindent \textbf{Acknowledgment}
This work was supported by the National Natural Science Foundation of China (Grants No. 11775159, No. 61621001 and No. 11674247), and Shanghai Pujiang Program under Grant No. 17PJ1409000, as well as the Fundamental Research Funds for the Central Universities.\\

\noindent \textbf{Author contributions}\\
W.W.Z., Y.S., Y.L. and H.C. performed the theoretical investigation; W.W.Z. carried out the COMSOL simulations; X. F. and D. L. designed and conducted the experiments; Y. L., Y. J., and H.C. guided the research; W.W.Z., Y.L., Y.J. and H.C. wrote the manuscript. All authors were involved in the analysis and discussion of the results. \\
\noindent \textbf{Competing financial interests}\\
The authors declare no competing financial interests.

\end{document}